\documentclass[useAMS,usenatbib]{mn2e}
\usepackage{amsmath,amssymb}                       

\usepackage{amsfonts}
\usepackage{graphicx}
\usepackage{float}
\usepackage{morefloats}
\usepackage{listings}
\usepackage{xcolor}

\usepackage{color,hyperref}
\definecolor{linkcolor}{rgb}{0,0,0.25}
\hypersetup{
  colorlinks=true,        
  linkcolor=linkcolor,    
  citecolor=linkcolor,    
  filecolor=linkcolor,    
  urlcolor=linkcolor      
}

\newcommand{\kms}{\ensuremath{\mathrm{km}~\mathrm{s}^{-1}}}

\definecolor{kvjcolor}{rgb}{0.75,0,0}

\definecolor{apwcolor}{rgb}{0,0.75,0.75}

\definecolor{jhcolor}{rgb}{0.901,0.46666,0.13333}

\definecolor{mpcolor}{rgb}{0,0.75,0.0}

\definecolor{kdcolor}{RGB}{150,0,250}

\definecolor{mwcolor}{rgb}{0.1,0.1,0.9}

\newcommand{\kmskpc}{\ensuremath{\mathrm{km}~\mathrm{s}^{-1}~\mathrm{kpc}^{-1}}}

\title[A bar's dark matter wake revealed by mSSA]
{The dark matter wake of a galactic bar revealed by multichannel Singular Spectral Analysis}
\author[J. A. S. Hunt et al.]
  {\parbox{\textwidth}{Jason A. S. Hunt$^{1}$\thanks{E-mail: j.a.hunt@surrey.ac.uk}, Michael S. Petersen$^2$, Martin D. Weinberg$^3$, Kathryn V. Johnston$^4$, Marcel Bernet$^{5,6,7}$, Kathryne J. Daniel$^8$, S\'oley \'O. Hyman$^8$, Adrian M. Price-Whelan$^9$, Arpit Arora$^{10}$ \& the EXP Collaboration}\vspace{0.5cm}
\\
$^1$ School of Mathematics \& Physics, University of Surrey, Stag Hill, Guildford, GU2 7XH, UK\\
$^2$ Institute for Astronomy, University of Edinburgh, Royal Observatory, Blackford Hill, Edinburgh EH9 3HJ, UK\\
$^3$ Department of Astronomy, University of Massachusetts at Amherst, 710 N. Pleasant St., Amherst, MA 01003\\
$^4$ Department of Astronomy, Columbia University, New York, NY 10027, USA\\
$^5$ Departament de Física Quàntica i Astrofísica, Universitat de Barcelona, C Martí i Franquès, 1, 08028 Barcelona, Spain\\
$^6$ Institut de Ciències del Cosmos, Universitat de Barcelona, C Martí i Franquès, 1, 08028 Barcelona, Spain\\
$^7$ Institut d’Estudis Espacials de Catalunya, Edifici RDIT, Campus UPC, 08860 Castelldefels, Spain\\
$^8$ Steward Observatory, University of Arizona, 933 North Cherry Avenue, Tucson, AZ, 85721, USA\\
$^{9}$ Center for Computational Astrophysics, Flatiron Institute, 162 5th Av., New York City, NY 10010, USA\\
$^{10}$ Department of Astronomy, University of Washington, Seattle, WA 98195, USA\\
}
\date{Accepted 2025 November 22. Received 2025 November 22; in original form 2025 October 10
}

\pagerange{\pageref{firstpage}--\pageref{lastpage}}
\pubyear{2022}

\begin{document}

\maketitle

\label{firstpage}

\begin{abstract}
The Milky Way is known to contain a stellar bar, as are a significant fraction of disc galaxies across the universe. Our understanding of bar evolution, both theoretically and through analysis of simulations indicates that bars both grow in amplitude and slow down over time through interaction and angular momentum exchange with the galaxy's dark matter halo. Understanding the physical mechanisms underlying this coupling requires modelling of the structural deformations to the potential that are mutually induced between components. In this work we use Basis Function Expansion (BFE) in combination with multichannel Singular Spectral Analysis (mSSA) as a non-parametric analysis tool to illustrate the coupling between the bar and the dark halo in a single high-resolution isolated barred disc galaxy simulation. We demonstrate the power of mSSA to extract and quantify explicitly coupled dynamical modes, determining growth rates, pattern speeds and phase lags for different stages of evolution of the stellar bar and the dark matter response. BFE \& mSSA together grant us the ability to explore the importance and physical mechanisms of bar-halo coupling, and other dynamically coupled structures across a wide range of dynamical environments.
\end{abstract}


\section{Introduction}
\setlength{\parskip}{0pt}
The Milky Way is known to contain a central bar since its presence was discovered through a combination of COBE photometry and the kinematics of gas in the inner galaxy \citep[e.g.][]{BS91,W92,W94,Detal95}. Since then, our view of the bar and the inner Galaxy have improved significantly with modern surveys \citep[see e.g.][for a summary]{Bland-Hawthorn:2016,Shen+20,Hunt+25}, and our understanding of bar dynamics and evolution has progressed through numerical simulation, and the study of barred galaxies throughout the universe.

For example, it has long been known that bars grow and slow through angular momentum transfer with the dark matter halo \citep[e.g.][]{Sellwood1980,Weinberg1985,Athanassoula2003}. Such a transfer affects not only the growth of the bar, but also the structure and velocity distribution of the dark halo. For example, simulations have shown the presence of a `shadow bar' counterpart to the stellar bar and a dark wake which extends further out into the halo \citep[e.g.][]{Petersen+16_shadowbar,Collier+Madigan2021,Ash+24,Marostica+2024,Frosst+2024}, and that spiral arms can also drive such a wake, albeit at a lower amplitude \citep{Bernet+25}. The \emph{shadow bar} is the dark-matter equivalent of the stellar bar which forms from the close-to-planar orbits of dark halo particles which are trapped into the Inner Lindblad resonance (ILR), just like the corresponding star particles \citep[see e.g.][for further discussion]{Petersen+16_shadowbar}.

\begin{figure*}
\centering
\includegraphics[trim={5cm 0 0 0},clip,width=\hsize]{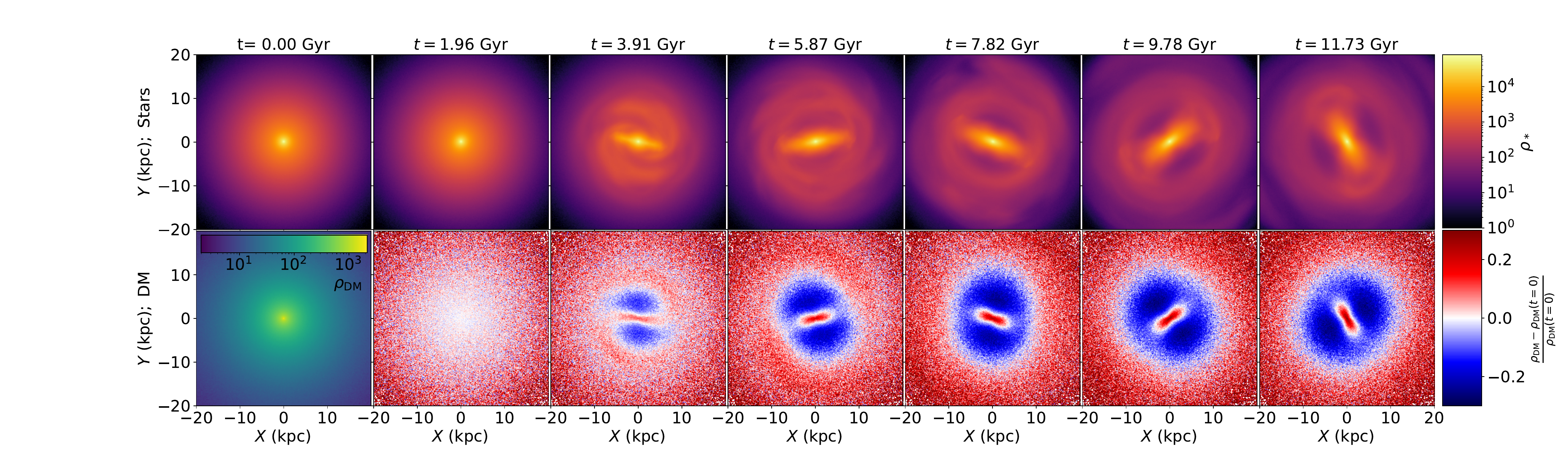}
\caption{\textbf{Upper row:} Face on number density of stellar particles as a function of time (increasing left to right). \textbf{Lower row:} Face on dark matter particle number density within $\mid z_{\mathrm{DM}}\mid<3$ kpc in the initial condition (left panel) and the relative number density over time compared to the initial condition (remaining panels) which show the evolution of the dark bar.} 
\label{fig:evo}
\end{figure*}

\begin{figure*}
\centering
\includegraphics[width=\hsize]{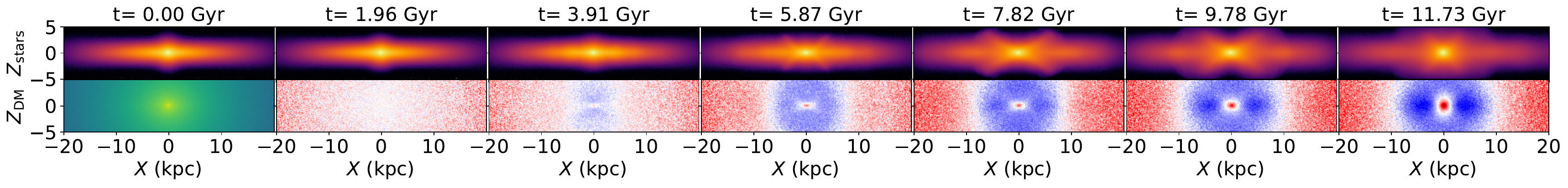}
\caption{\textbf{Upper row:} Edge on number density of stellar particles as a function of time (increasing left to right). \textbf{Lower row:} Edge on dark matter particle number density in the initial condition (left panel) and the relative number density over time compared to the initial condition (remaining panels) which show the evolution of the dark bar. The color bar is the same as Figure \ref{fig:evo}.} 
\label{fig:edge}
\end{figure*}

However, it can be difficult to separate and quantify such dynamical processes within live $N$-body simulations owing to the complexity of the representation, and the combination of subtle overlapping modes. Understanding the physical mechanisms underlying this coupling requires modelling of the non-axisymmetric time evolution of the disc and halo simultaneously.

A Basis Function Expansion (BFE) of the simulation is a powerful way to generate a compressed representation of the entire model evolution. Using a global BFE \citep[e.g.][]{CB72,Kalnajs76,Weinberg89} we can generate a time series of coefficients that capture the evolution of galactic structure. By choosing an adaptive basis with global support \citep[e.g. the method from][which captures both spheroid and disc components]{Weinberg99} where the simulation is represented in a small number of coefficients, we obtain a drastically compressed non-parametric representation of the simulation.

We can then apply multichannel Singular Spectral Analysis \citep[mSSA; e.g.][]{Ghil+Vautard1991,Golyandina+02} to these coefficient series in order to identify the spatial and temporal correlations which represent the dynamical modes present in the simulation. This is a relatively recent innovation which addresses the above challenges; namely that mSSA can explicitly extract and quantify dynamical modes from within large and complex particle-based simulations when coupled with a Basis Function Expansion coefficient representation of such a simulation \citep[e.g.][]{Weinberg+21,Johnson+23,Arora+2025}. 

\cite{Weinberg+21} used BFE and mSSA to analyse the evolution of a barred stellar disc, and \cite{Johnson+23} used BFE and mSSA to analyse disc-halo coupling in an unbarred disc-halo system. In this work, we build upon their work to both illustrate and quantify the explicitly coupled evolution of the stellar bar and dark matter halo, including the growth of the dark bar, and the dark wake. In Section \ref{sec:barsim} we describe the barred galaxy simulation used in this work. In Section \ref{sec:coef-gen} we describe how we post-process the simulation into the BFE coefficients necessary for the subsequent analysis, and show how they can be used to estimate the bar pattern speed in Section \ref{sec:patternspeed}. In Section \ref{sec:mssa_method} we briefly describe the mSSA methodology as applied to galaxy simulations \citep[see][for a more detailed description]{Weinberg+21,Johnson+23}, and in Section \ref{sec:mssa_pc} we describe the extraction of Principal Components. In Section \ref{sec:mssa_apply} we demonstrate the use of the method to extract the coupled dynamics of the bar and the dark matter halo including the coupled quadrupole, and in Section \ref{sec:mssa_candc} we demonstrate causation, including differences in growth rates and phase differences. In Section \ref{sec:summary} we give our summary and conclusions.


\section{Simulation \& Methodology}\label{sec:method_and_sim}

In order to explore the coupling of the stellar bar and the dark matter halo, we perform a high-resolution isolated $N$-body simulation of a barred disc galaxy as described in Section \ref{sec:barsim}. We then post-process it using basis function expansion as described in Section \ref{sec:coef-gen}.

\subsection{Simulation}\label{sec:barsim}
For our fiducial barred galaxy simulation we make use of the isolated host N-body Milky Way Galaxy model, [MW], from \cite{Stelea+24}. We evolve the galaxy for a further 9 Gyr for a total of $\sim12$ Gyr of evolution using the $N$-body tree code \texttt{Bonsai}\rm\ \citep{Bonsai,Bonsai-242bil}. We use an opening angle of $\theta_{\mathrm{o}}=0.4$ rad, and a softening length of 10 pc.

For full details of the simulation setup see \cite{Stelea+24}, but in brief; the initial conditions for the galaxy were generated using the galactic dynamics software package \texttt{Agama}\ \citep{agama} based on the Milky Way-like host galaxy with an axisymmetric halo from \cite{Vasiliev+21}. The galaxy consists of a $1.2\times10^{10}\ M_\odot$ spherical bulge and a $7.3\times10^{11}\ M_\odot$ dark halo with $8\times10^7$ and $6\times10^8$ particles respectively. The bulge and halo follow \texttt{Agama}'s\rm\ Spheroid potential;
\begin{equation}
    \rho=\rho_0\biggl(\frac{\tilde{r}}{R_\mathrm{s}}\biggr)^{-\gamma}\biggl[1+\biggl(\frac{\tilde{r}}{R_\mathrm{s}}\biggr)^{\alpha}\biggr]^{\frac{\gamma-\beta}{\alpha}}\times\exp\biggl[-\biggl(\frac{R}{R_{\mathrm{cut}}}\biggr)^{\xi}\biggr],
    \label{eqn:spheroid}
\end{equation}
where $R_\mathrm{s}=0.2,7.0$, $\alpha=1,2$, $\beta=1.8,2.5$ and $\gamma=0,1$ for the bulge and halo respectively. The $5\times10^{10}\ M_\odot$ disc contains $3.2\times10^8$ particles and follows an exponential profile with $R_\mathrm{d}=3$ kpc, $h_\mathrm{d}=0.4$ kpc and $\sigma_{R_0}=90$ \kms. As described in \cite{Stelea+24}, this creates a disc which is relatively warm, and it takes $\sim2$ Gyr before a bar becomes visually apparent owing to the overall stability of the initial condition. As such, this model is not meant to precisely reproduce the Milky Way, but instead provides us a relatively well behaved laboratory on which to test our method. We defer a systematic study of bar growth as a function of such disc and halo properties to future work.

The upper row of Figure \ref{fig:evo} shows the  face on density evolution of the disc over time (increasing left to right panels). The bar grows over time from an initially smooth disc to a very strong bar at late times. The final state of the bar in the right-hand panels is significantly stronger than the bar of the Milky Way, and the `Milky Way-like' stage is estimated to be approximately $t\sim3-4$ Gyr. In this work we analyse the full 11.7 Gyr of evolution beyond the Milky Way-like stage, as the goal of this paper is to illustrate the power of the tools in their ability to reveal coupled dynamical modes in the disc and halo, not to reproduce the Milky Way's bar specifically.

The lower row of Figure \ref{fig:evo} shows the evolution of the dark matter density in a slice where $\mid z\mid\leqslant5$ kpc from the midplane. The left-most panel shows the number density, while the remaining panels show the fractional difference in density compared to the left panel. There is a clear overdense `dark bar' component in the inner galaxy which qualitatively matches the angle of the stellar bar in the upper row. There is also a general rearrangement of the halo particles to become less concentrated, developing a relative underdensity within $\sim10$ kpc, and overdensity outside this radius. We also note here that the dark matter response to the stellar spiral arms is faintly visible (on the order of a few percent) in the third and fourth columns, as described in \cite{Bernet+25}. We choose here to focus on the bar-halo coupling, and defer exploration of the spiral arm growth to future work.

Figure \ref{fig:edge} shows the equivalent figure but for the edge on projection. The top row shows that a strong X feature evolves over time. We note here that in this model the X grows resonantly and the bar does not undergo a buckling event. The lower row shows the dark matter only appears to be trapped in the planar bar, with no obvious counterpart to the X, instead showing again the radial rearrangement of the dark matter away from the inner $\sim10$ kpc. We also note that at later times the dark matter halo undergoes the dipole instability described in \cite{Weinberg_2023_dipole_instability}, but it does not affect the subsequent analysis or conclusions in this work.

\begin{figure*}
\centering
\includegraphics[width=\hsize]{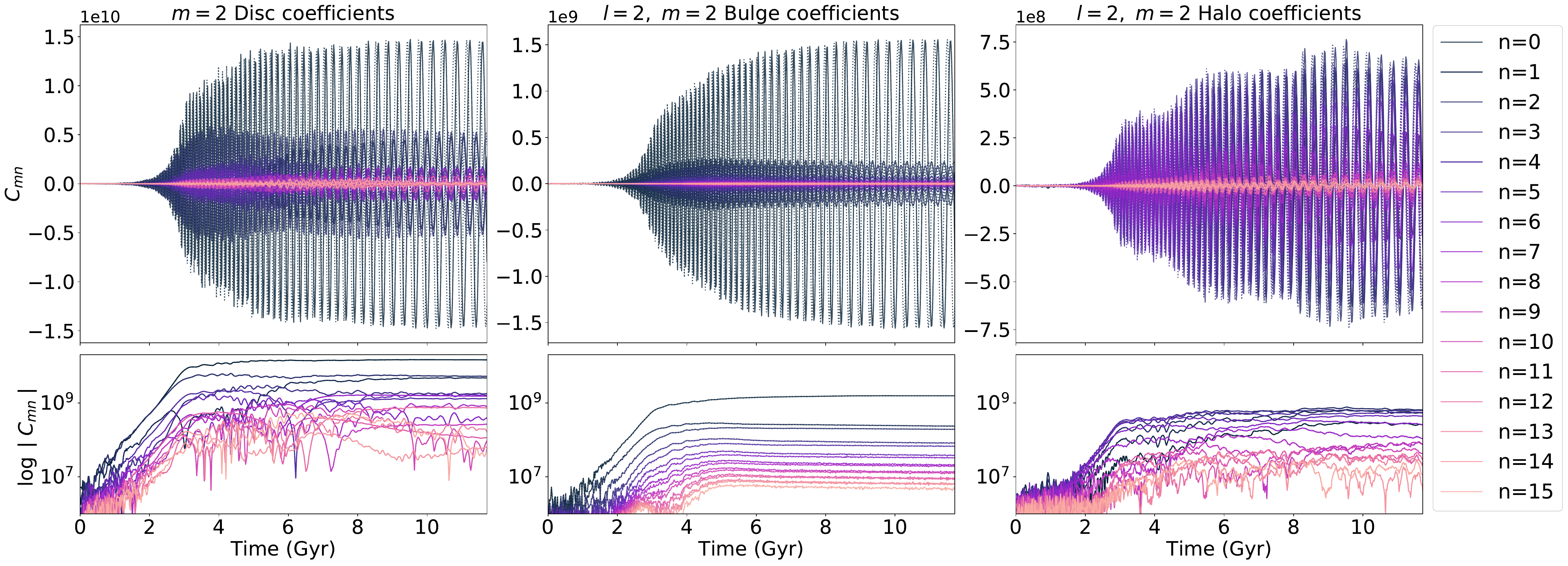}
\caption{\textbf{Upper row:} Basis function expansion coefficients for $m=2,\ n=0-15$ for the vertically symmetric disc basis functions (left) and  $l=2,\ m=2,\ n=0-15$ for the spherical basis functions representing the stellar bulge (middle) and dark matter halo (right). The radial scale increases with decreasing values for $n$, i.e. low $n$ represents the largest scales. The $m=2$ disc and bulge coefficients primarily show the growth and subsequent evolution of the bar, while the halo coefficients show the response of the dark matter halo. \textbf{Lower row:} Absolute value of the coefficients showing the expected exponential growth in the bar instability at early times, while losing phase and pattern speed information.}
\label{fig:coefs}
\end{figure*}

\subsection{Representation of simulation in BFE coefficient time series, $A_{nlm}$}\label{sec:coef-gen}

The previous studies of bar evolution with basis function expansion of \cite{Petersen+19,Weinberg+21,PWK21a} were performed using a simulation run with the BFE based simulation code \texttt{exp}\ \rm \citep{PWK21,exp2025JOSS}, which naturally produces the coefficient time series as a product of the force evaluation. 

In contrast, the simulation which is analysed in the work was performed with the N-body tree code, \texttt{Bonsai}\rm. Thus, we first post-process the simulation shown above to calculate the basis function coefficients needed for our mSSA analysis as previously described in \citep{Johnson+23} although they used a two dimensional basis for the disc which differs from the three dimensional disc basis used in this work. The steps are as follows, and tutorials can be found online\footnote{\url{https://github.com/EXP-code}} along with the publicly available simulation and analysis framework \texttt{exp}\ \citep{exp2025JOSS}. 

Firstly, we must construct a basis for each galactic component. Our $N$-body galaxy is made up of two spherical components (a stellar bulge and a dark matter halo) and a single stellar disc component, as described in Section \ref{sec:barsim}. 

For the stellar bulge and the dark matter halo we construct our basis by matching the lowest-order basis function pair to the model through an eigenanalysis solution to the Poisson equation as detailed in \cite{Weinberg99} and updated in \cite{PWK21}, which generates a spherical basis characterised by standard spherical harmonic indices $l,m,n$, where these indices correspond to the number of nodes in the basis functions. For the construction itself, we start by calculating a spherically symmetric radial density profile from the particle distribution in the initial condition of the bulge and the dark halo\footnote{Note that we could also straightforwardly provide an analytic profile based on the expected distribution from equation \ref{eqn:spheroid}, but calculating it directly from the particle distribution generalizes to more complex systems}. 

We then construct the basis such that the lowest order function matches the density profile of the initial condition. The higher order basis functions are then generated as eigenfunctions of the Sturm-Liouville equation conditioned on the potential-density model. For the dark halo we calculate the density profile in 1000 logarithmically spaced bins from $0.0007-600$ kpc, with maximum harmonics of $l_{\mathrm{max}}=10$ and $n_{\mathrm{max}}=48$. The basis construction also requires a scale parameter which we set as $R_{\mathrm{mapping}}=7$ kpc to match the scale radius of the dark halo. For the bulge, we repeat the process for 350 bins from $0.1-7$ kpc, and set $R_{\mathrm{mapping}}=0.2$ kpc, for $l_{\mathrm{max}}=10$ and $n_{\mathrm{max}}=32$ \citep[see also][for an example applying a such a spherical expansion to dark matter halos from the FIRE simulations]{Arora+2025}.

For the disc, it is technically possible to construct such a spherical basis which can reproduce a flat disc-like component if the basis is constructed with a sufficiently high $l_{\mathrm{max}}$. However, in practice this is inefficient and inaccurate.  We instead construct a basis of cylindrical basis functions which better represent a disc-like distribution. 

We use the `empirical orthogonal function' (EOF) approach as is described in full in \cite{Weinberg+21,PWK21}, but in brief; one can find an orthonormal transformation between a `high-order' spherical basis and a `low-order' cylindrical basis which accurately represents the disc. Similar to the spherical case we condition the basis such that the lowest order function represents the initial analytic mass distribution\footnote{although it is not guaranteed to be exact in the EOF approach.}, and the higher order functions are modifiers on top of this monopole. 

Unlike in the spherical case we do not have $l,m,n$, but instead compose the basis of multiple `families' of cylindrical basis functions, namely those which are vertically symmetric, those which are vertically antisymmetric, and those which describe vertical compression. For an isolated disc with little vertical distortion the vertically symmetric functions are the most important, but the antisymmetric basis functions are essential for describing departures from vertical equilibrium, such as in satellite-disc interactions \citep[e.g. see][for a `BFE based' analysis of this galaxy interacting with Sgr and the LMC]{Petersen+25} 

In this instance we set the disc type  $d_{\mathrm{type}}=\mathrm{exponential}$ and the parameter $a_{\mathrm{cyl}}=3.0$ kpc and the scale height, $h_\mathrm{cyl}=0.8$ kpc to match the simulation disc scale length and height,. We set the maximum azimuthal harmonic $m_{\mathrm{max}}=6$, and for the radial harmonics we set $n_{\mathrm{max}}=72$. We set the number of antisymmetric functions $n_\mathrm{odd}=36$, such that there are 36 symmetric functions ($n_{\mathrm{max}}=n_{\mathrm{even}}+n_{\mathrm{odd}}$). {We also note here that we repeated the subsequent analysis with a different `thinner' disc basis (with $h_{\mathrm{cyl}}=0.28$ kpc, $n_{\mathrm{max}}=30$ and $n_\mathrm{odd}=12$, with the other parameters remaining the same) and found the same dynamical modes and conclusions. As such, the dynamical analysis in this work is not sensitive to the choice of disc basis.

Once we have these bases, we can calculate the coefficients which correctly weight our sets of basis functions to reproduce the particle data in each simulation snapshot. We again here use \texttt{pyEXP}\rm\ to calculate the coefficients for each of the three components (i.e. disc, bulge and halo). The coefficients for each simulation snapshot are independent and can be calculated in parallel before combining them into the full time series.

\section{Raw coefficient analysis}\label{sec:coefs}

By construction, the basis coefficients themselves encode the structural properties of the disc and halo. In this work, which focuses on the dynamics of the bar, we examine the harmonics which are related to bar-like structure, e.g. for the disc we investigate the azimuthal harmonic $m=2$ corresponding to the quadrupolar bar structure. For the stellar bulge and dark halo we investigate the corresponding spherical harmonics, $l=2$ (corresponding to the vertically `squashed' disc-like component), and azimuthal $m=2$. 

Although a simple bar model may be purely quadrupolar, in reality bars are commonly composed of a combination of $m=2$ and higher-order even harmonics in both simulation \citep[e.g.][]{Hunt+18_41OLR} and observation \citep[e.g.][]{Buta+06}. However, in this initial work we choose to focus upon the dominant quadrupole term and defer exploration of the higher-order modes to future work. 

\subsection{Overview of disc, bulge and halo coefficients}\label{sec:coef_overview}

The upper left panel of Figure \ref{fig:coefs} shows the raw $m=2$ disc coefficients for the sixteen vertically symmetric radial harmonics $n=0-15$, for the sine (dotted) and cosine (solid) components, while the lower left panel shows the absolute amplitude of the respective series of coefficients. Note that $n$ corresponds to the number of radial nodes in the basis function, and thus increasing $n$ corresponds to smaller radial scales. Thus $n=0$ is the largest `disc wide' signature. The growth of the bar is immediately clear from early times in all $n$. There are also clearly different evolutionary regimes that are more apparent in the higher $n$'s. The underlying dynamical meaning of these coefficients and the distinct dynamical regimes are discussed in Section \ref{sec:mssa}, but for now we use the coefficients purely to show that the different phases of bar evolution are immediately apparent in the disc, bulge and halo $m=2$ coefficients. 

The central column of Figure \ref{fig:coefs} shows the raw $l=2,\ m=2$ bulge coefficients for the first sixteen radial harmonics for the sin (dotted) and cos (solid) components. Again, these lower $n$'s correspond to larger scale radial structure in the bulge. The bulge $l=2$, $m=2$ mode grows with the disc. It is very smooth, and shows no interesting features in the later evolution, thus we exclude it from the subsequent analysis which focuses on bar--halo coupling.

\begin{figure}
\centering
\includegraphics[width=\hsize]{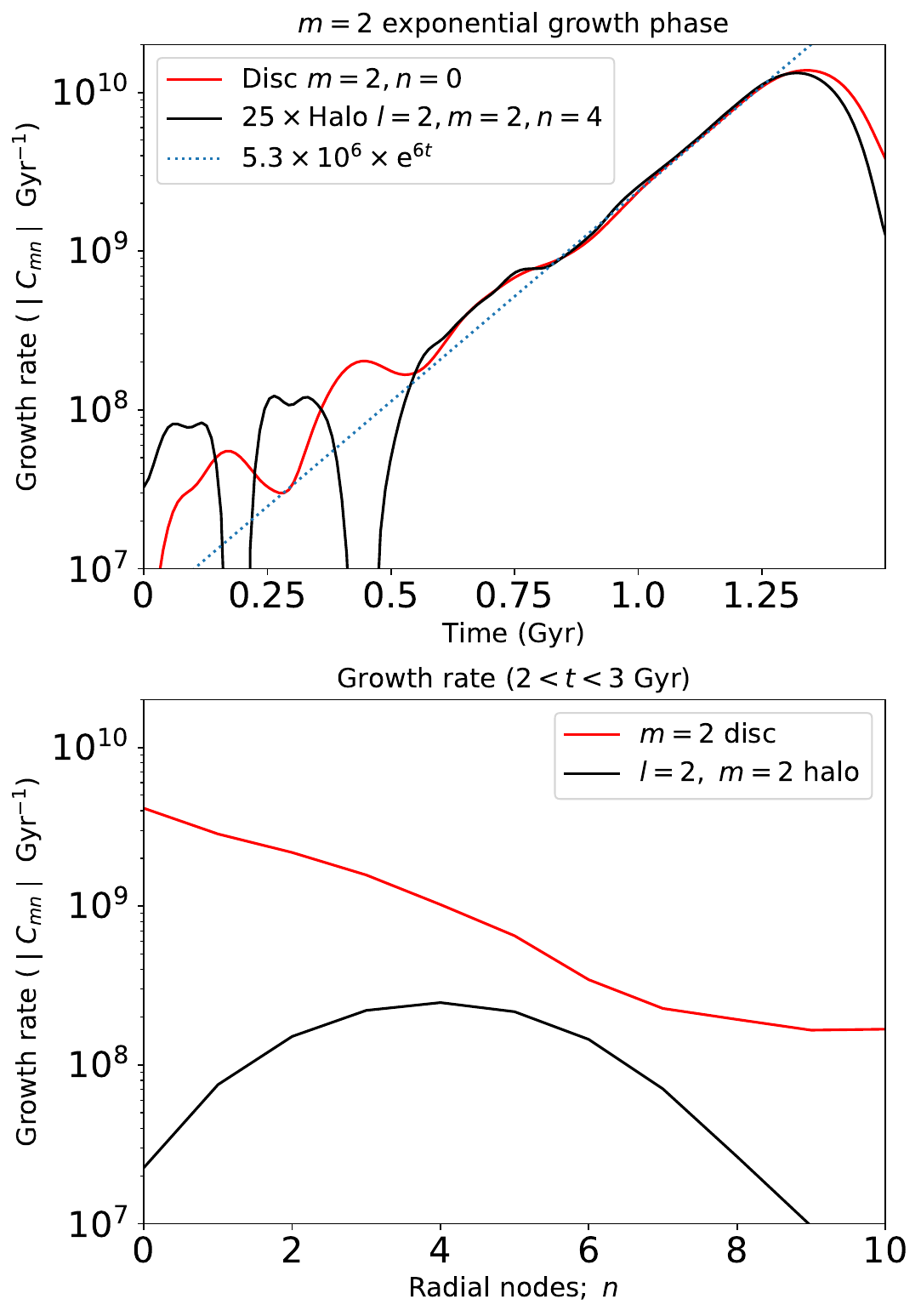}
\caption{\textbf{Upper:}\ Growth rate of the $m=2$, $n=0$ coefficients for the disc (red) and the $l=2,\ m=2$, $n=2$ coefficients for the halo (black) as a function of time during the early formation of the bar. The growth rate at these times increases exponentially with $\mathrm{e}^{6t}$. \textbf{Lower:} Growth rate from $2\leqslant t\leqslant3$ Gyr, for radial nodes $n=0-10$. Growth rates are defined as the change in coefficient amplitudes over time.
}
\label{fig:bar_growth}
\end{figure}

\begin{figure*}
\centering
\includegraphics[width=\hsize]{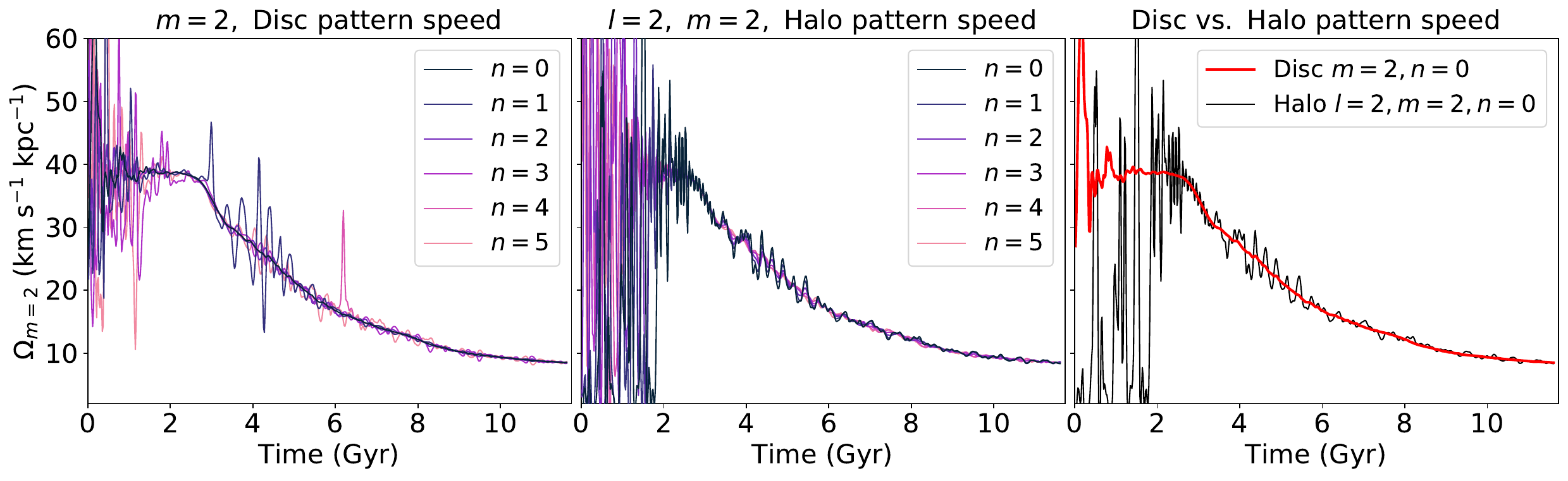}
\caption{Pattern speed of the disc $m=2,\ n=0-5$ coefficient series (left) and the halo $l=2,\ m=2,\ n=0-5$ coefficient series (middle) as shown in Figure \ref{fig:coefs}. The right panel shows the disc $m=2,\ n=0$ series vs. the halo $l=2,\ m=2,\ n=0$ series, which are clearly coupled. 
}
\label{fig:patternspeeds}
\end{figure*}

The right column of Figure \ref{fig:coefs} shows the raw $l=2,\ m=2$ coefficients for the first sixteen radial harmonics for the dark matter halo. In contrast to the disc and bulge coefficients, the halo signal appears to grow slightly later than in the disc, at $t\sim2$ Gyr. This is as we would expect if the bar evolution is the driving influence on the halo dynamics, although at this stage this is merely a qualitative correlation `by eye'. We also note that while the bar and bulge signal was strongest in the $n=0$ coefficient series, the response in the halo appears to be stronger at slightly higher $n$'s than at $n=0$ (with highest amplitude for $n=2$). This is to be expected as the radial extent of the disc (and bulge) is much smaller than that of the halo, and thus the largest disc scales will correspond to intermediate halo scales.

Regardless, multiple phases of the halo evolution are immediately clear again in the halo coefficients. There is some exponential growth phase from $2\lesssim t\lesssim3$ Gyr, followed by some intermediate slower growth phase from $3\lesssim t\lesssim5$ Gyr, and an approximately steady state for $5\lesssim t \lesssim 8$ Gyr. At this point there is a transition to another apparently stable state which persists for the remainder of the simulation.

\subsection{Indications of coupling}\label{sec:coupling}
\subsubsection{Growth rates}\label{sec:growth_rates}

The upper panel of Figure \ref{fig:bar_growth} shows the growth rate, defined as the change in coefficient amplitude over time, of the $m=2$, $m=0$ disc and $l=2,\ m=2$, $n=2$ halo coefficients (i.e. the highest amplitude $n$ order for both coefficient series) from $0<t<1.4$ Gyr. Both the disc and halo modes show exponential growth during the bar formation phase, as expected from linear theory \citep[e.g.][]{Petersen+24}, following $\sim\mathrm{e}^{6t}$. This indicates that the $m=2$ instability is present from the start of the simulation, and as such we do not find a well defined bar formation time. Figure \ref{fig:bar_growth} also shows that the shadow bar initially grows at the same rate as the stellar bar, before the resonant coupling to the halo kicks in. 

The lower panel of Figure \ref{fig:bar_growth} shows the growth rate in the latter non-exponential growth phase from $2<t<3$ Gyr which was apparent in the coefficients from Figure \ref{fig:coefs} as a function of radial orders $n=0-10$. The $m=2$ mode grows at different rates for different radial orders. Note in particular that while the $m=2$ disc instability grows fastest for low $n$, the halo $l=2,\ m=2$ mode grows fastest for intermediate $n$'s, again owing to the different scales represented by $n$ in the disc and halo. 

\subsubsection{The Pattern Speed}\label{sec:patternspeed}

For both the disc and the halo coefficients in Figure \ref{fig:coefs}, the frequency of the pattern visually decreases with time, corresponding to the slowdown of the bar. We can also directly measure the pattern speed of the bar and halo response by using the real and imaginary components of the coefficient time series to compute the position angle, and how it changes over time. Figure \ref{fig:patternspeeds} shows the pattern speed of the series of disc coefficients $m=2,\ n=0-5$ (left), the series of halo coefficients $l=2,\ m=2,\ n=0-5$ (middle) and the disc series $m=2,\ n=0$ vs. the halo series $l=2,\ m=2,\ n=0$ (right).

The $m=2$ pattern speed shown in the left panel corresponds to the bar. On the largest scales ($n=0$) the $m=2$ pattern has a coherent frequency just under $t=1$ Gyr, and remains constant ($\Omega_{\mathrm{p}}\sim39$ km s$^{-1}$ kpc$^{-1}$) until $t\sim2$ Gyr. This shows that the $m=2$ instability is present from a time earlier than can be seen in the coefficient amplitudes in Figure \ref{fig:coefs}, but covers the period of exponential growth in the upper panel of Figure \ref{fig:bar_growth}. From $t\sim2$ Gyr the $m=2$ disc component starts to slow down. The middle panel of Figure \ref{fig:patternspeeds} shows that the pattern speed of the $l=2,\ m=2$ component of the halo is visually consistent with noise until $t\sim$ 2 Gyr, at which point the halo $l=2,\ m=2$ pattern appears to slow at the same rate as the disc $m=2$.

The right panel shows the same lines for the $n=0$ coefficient series from the left and middle panels, which are clearly consistent. Notably, while the disc $m=2$ mode exists at $\lesssim2$ Gyr, the disc $m=2$ mode starts to slow down only once it couples with the halo at $t\sim$ 2 Gyr, the same time at which it starts to grow in amplitude. This is consistent with expectations from disc halo coupling and the transfer of angular momentum which have long been known in the literature, but the behavior is clearly illustrated through the use of basis function expansion. 

However, at this stage (i.e. Section 3) our interpretation is driven by a \textit{qualitative} `by eye' comparison which finds dynamical behavior which is already expected based on known dynamics. In addition, while the $m=2$ signal is clearly dominated by the bar, these coefficients will capture all $m=2$ patterns such as the spiral structure (although we would expect spiral arms to primarily present at higher $n$'s; i.e. at finer radial scales). In order to decompose the individual dynamical signatures, we use multichannel Singular Spectral Analysis as described in the next section.

\begin{figure*}
\centering
\includegraphics[width=\hsize]{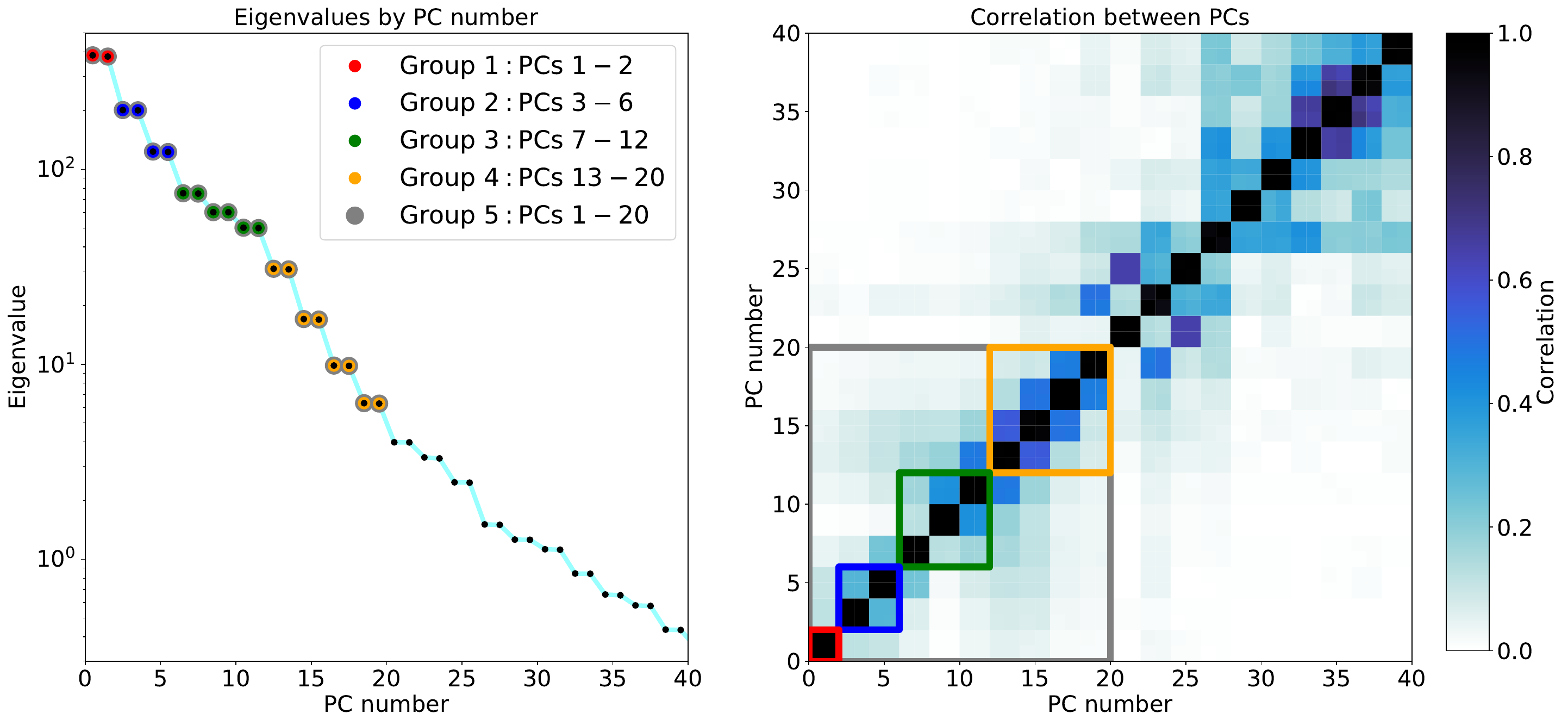}
\caption{Eigenvalues by PC number (left column) and correlation between PCs (right column), for the mSSA correlation of the disc $m=2,\ n=0-7$ and halo $l=2,\ m=2,\ n=0-7$ coefficient time series using a window of $L=40$. The PCs are colored by their assigned grouping, as discussed in the text. The darker bands of off diagonal power indicate incomplete separation, or over-separation of dynamical signals as discussed in the text.}

\label{fig:LowLcorrelationmatrix}
\end{figure*}

\section{Identifying dynamical structure with multichannel Singular Spectral Analysis}\label{sec:mssa}

\subsection{mSSA}\label{sec:mssa_method}
\subsubsection{Methodology}

Singular Spectral Analysis \citep[SSA; e.g.][]{Ghil+Vautard1991,Golyandina+02} is a method which can decompose some time series into a sum of some individual signals correlated by dynamical variation. The correlation is computed by performing Principal Component Analysis (PCA) within a `trajectory matrix' which consists of sequentially lagged, windowed versions of the time series, such that SSA can find temporal correlations within the series. Multichannel Singular Spectral Analysis (mSSA) extends SSA to be able to take in and cross-correlate multiple time series, identifying signals which are present in multiple series.

In our application of mSSA, our time series are the BFE expansion coefficients discussed in the previous section. By using the multichannel version, we can correlate multiple coefficient series against each other. For example, we can correlate several harmonic orders against each other, or correlate series from different galactic components, e.g. in our case, the disc and dark halo. A detailed explanation of mSSA along with example decompositions can be found in \cite{Weinberg+21} and appendix C of \cite{Johnson+23}. An application of mSSA to disentangle the impact of a Sgr-like dwarf and an LMC-like dwarf impacting this host galaxy can be found in the companion paper \cite{Petersen+25}, and a recent application to the dark matter halos of cosmological simulations can be found in \cite{Arora+2025}.

In this section we demonstrate the power of mSSA to separate unrelated dynamical signals from each other, and to find correlated dynamical evolution in the disc and halo. 

\subsubsection{mSSA options}\label{sec:mssa_choices}
The choices of both which harmonics to feed in to mSSA, and the choice of window size, $L$, for the cross correlation will determine the dynamical modes which the analysis is most sensitive to (where $L$ is effectively the number of snapshots being correlated at a time). In this work, we feed in the disc $m=2$ coefficient series and the halo $l=2,\ m=2$ coefficient series for the first eight radial harmonics, $n=0$ to 7. This will then be most sensitive to the large radial scale $m=2$ modes in the disc and halo, such as the bar and dark matter wake, while being less likely to pick up fine features such as material spiral arms, which we would expect to present on finer radial scales, i.e. at higher $n$.

The window length, $L$, is an imposed trade off of low-frequency ambiguity with high-frequency insensitivity. For example, if we make the $L$ as large as possible (half the length of the time series, or $L=300$ in this case for our 600 snapshots) we run the risk of artificially breaking up a slowly varying long-term signal with changing frequency into chunks with different frequencies. This is an artifact of the multi-channel nature of mSSA that would not occur with a single channel. To be explicit: suppose there is only one quasi-periodic signal spread over all the channels. In the presence of noise, one can get a better fit to multiple frequencies over shorter time domains than a single signal with a slowly changing frequency. One can improve this tendency to artificially chop up a single signal into multiple regimes by using shorter temporal segments in the trajectory matrix. This can help to make a physical interpretation, as shown in \cite{Weinberg+21}, and in this work.

As such, we set the window length to $L=40$ in this work. However, we note that the results in the below sections are not sensitive to the exact choice of $L$, providing $L\ll300$ and above some minimum threshold to recover periodic signals.

\begin{figure}
\centering
\includegraphics[width=\hsize]{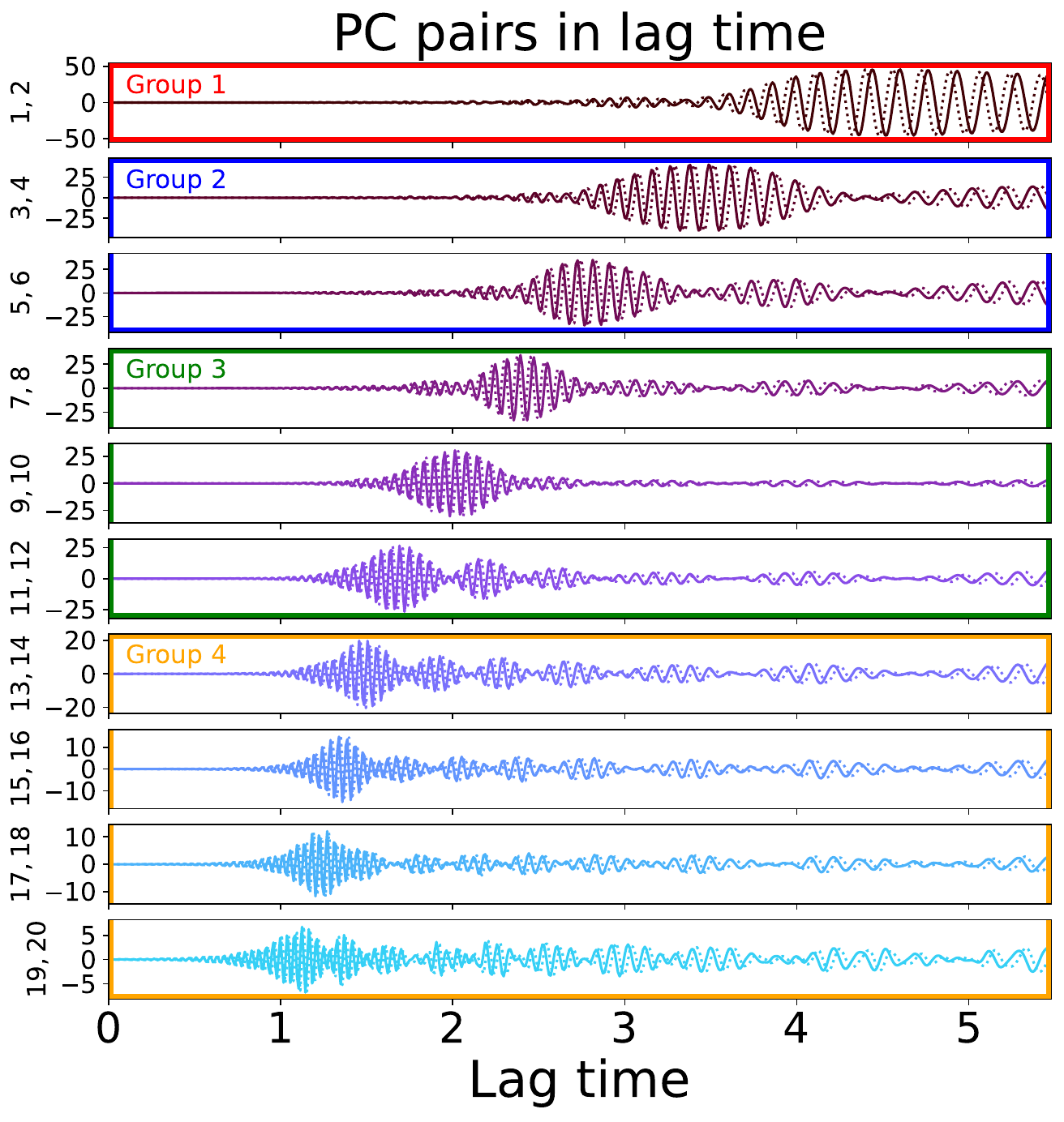}
\caption{The first ten PC pairs in `lag time' for $L=40$, matching those of Figure \ref{fig:LowLcorrelationmatrix}. The PC pairs are visually consistent within the pair, and distinct from other pairs. Note that the higher PCs (with lower significance) occur earlier in lag time, with some overlap. The groups are illustrated with colored axis frames matching other Figures.}
\label{fig:LowLPCs}
\end{figure}

\begin{figure}
\centering
\includegraphics[width=\hsize]{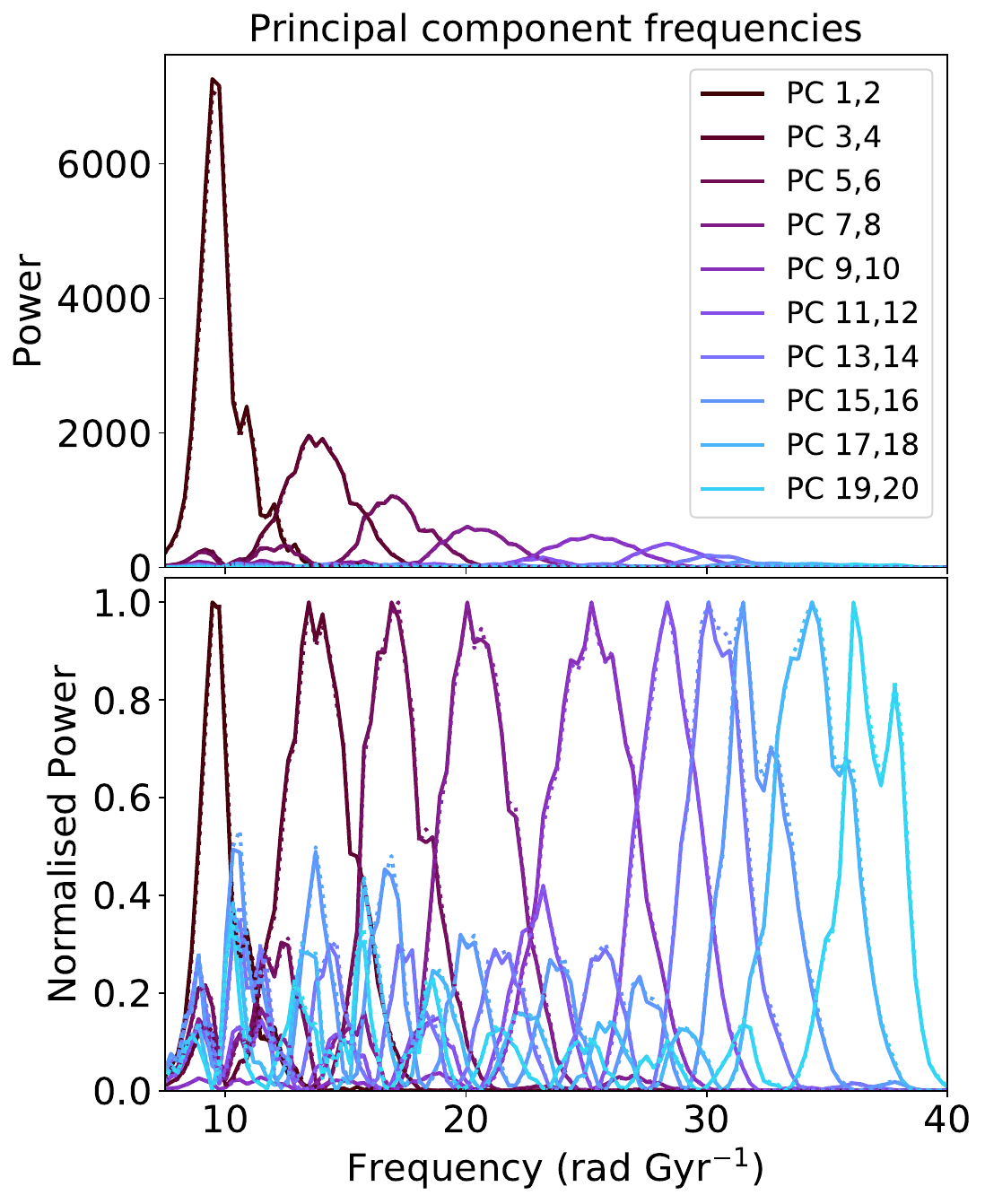}
\caption{Power (upper) and normalised power (lower) at different frequencies for the first 20 PCs matching those shown in Figure \ref{fig:LowLPCs}. The PCs are paired (even dotted, odd solid lines) as seen in the eigenvalues and the correlation matrix of Figure \ref{fig:LowLcorrelationmatrix}.}
\label{fig:LowLfreqs}
\end{figure}

\subsection{Principal components \& grouping}
\label{sec:mssa_pc}
\subsubsection{\bf Considering the PC's}

Performing mSSA on the coefficient series with the above options produces a set of principal components (PCs) which represent correlated signals in the disc and halo coefficient time series.

Figure \ref{fig:LowLcorrelationmatrix} shows the eigenvalues of the PCs, ordered by their amplitude (left column) and the correlation matrix (right column) for the mSSA decomposition with $L=40$. Note that it is expected for the PCs to come in pairs for quasi-periodic signals, which encodes phase information. This can be seen both in the eigenvalues in the left column where there are many pairs of PCs with similar significance, and also in the correlation matrix in the right column where PC 1 \& 2 are perfectly correlated, as are 3 \& 4 etc, leading to the diagonal. The eigenvalues of the PCs are colored by their grouping as discussed in the next section, and the same groups are marked with boxes on the correlation matrix. 

There are also significant non-diagonal terms in the correlation matrix, implying imperfect separation of the underlying dynamical modes. In fact, in this instance we know that the PCs are in fact `over-separated' in that the small window length has induced artificial separation of the `bar' signature into distinct temporal regimes. These distinct regimes may be dynamically distinct, as explored below.

The upper set of Figure \ref{fig:LowLPCs} shows the first ten PC pairs in `lag time' as defined below. The PC pairs are visually consistent within each pair, yet they are distinct from other pairs, showing a clear and consistent shift towards lower `lag time' with increasing PC number.

\begin{figure*}
\centering
\includegraphics[width=\hsize]{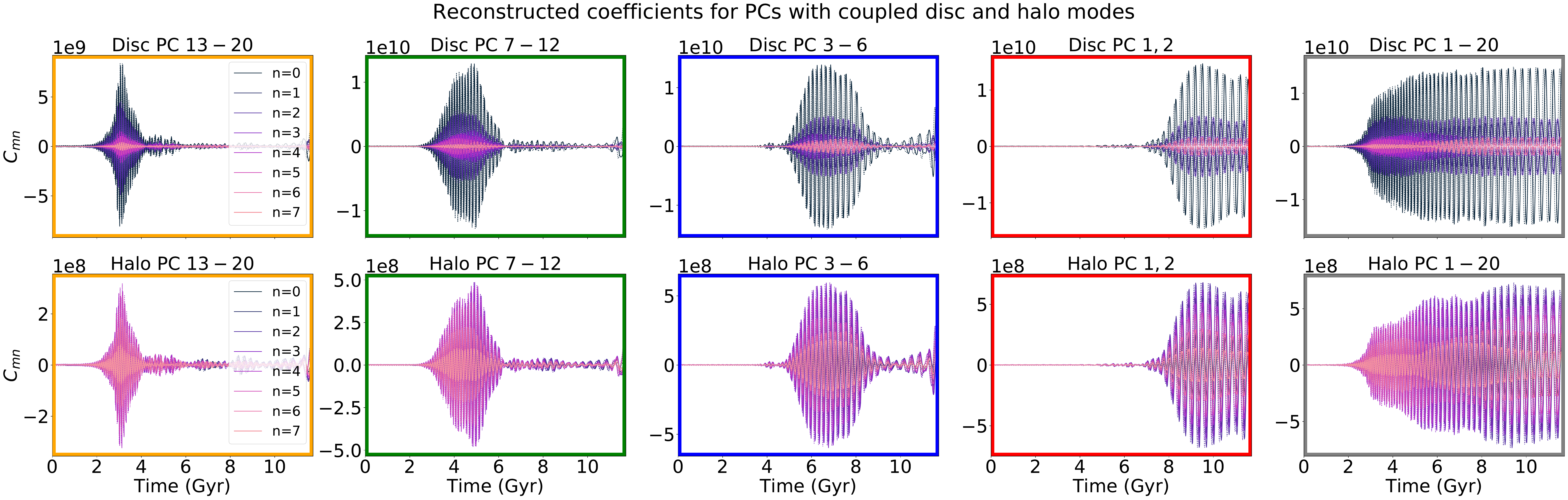}
\caption{Reconstructed $m=2,\ n=0-7$ disc coefficients (upper row) and $l=2,\ m=2,\ n=0-7$ halo coefficients (lower row) for the PC groups which broadly correspond to different evolutionary phases of the bar, as discussed in the text (left four columns), and the combination of all 20 PC pairs (right column) showing the full reconstruction of the $m=2$ signal matching Figure \ref{fig:coefs}. Note the excellent correlation between disc and halo.}
\label{fig:LowLreconCoefs}
\end{figure*}

`Lag time' is not `physical time' and there is not a direct mapping to the simulation time. Lag time has an indistinct zero point because the value of a PC at some lag time come from many physical times. To be explicit, the run of time in some PC is the same as the run of time in the physical simulation. However, the signal making up a particular PC comes from pieces of the time series that have been shifted or `lagged' in physical time. So the PC does not directly correspond to a signal in physical time. This projection of the PCs in `lag time' is informative, but we also wish to investigate the impact of each dynamical PC on the dynamical evolution of the disc in `physical' time. For this, we must reconstruct the contribution of the PCs to the original time series of coefficients, as done below in Section \ref{sec:recon_coefs}.

However, while lag-time and physical time are not equivalent, the time steps in lag time and physical time are the same. Thus, we can perform a discrete Fourier transform of a PC and learn something precise about the physical frequency (i.e. pattern speed) of the dynamics represented by that PC. Figure \ref{fig:LowLfreqs} shows the power vs. frequency for the first ten PC pairs (upper) and the power normalised to the peak of each PCs power (lower) to illustrate the shift in frequencies. We remind the reader that this is \emph{variance} power which depends on the mSSA detrending scheme as discussed in \cite{Weinberg+21}.

The dotted and solid lines in Figure \ref{fig:LowLfreqs} represent the odd and even PC respectively. Each PC pair occupies a small section of frequency space, with some small overlap between them. There is a general trend of decreasing power with increasing frequency, i.e. the $m=2$ component of the bar grows stronger as it slows down, as expected. We also note that the order of PC pairs appears to increase by frequency, yet this is not required to be the case, with the ordering instead depending on the eigenvalues (see Figure \ref{fig:LowLcorrelationmatrix}). The reason why PCs appear ordered in frequency is that the bar spends more of its duration at lower frequencies. 

The lower panel shows that the PCs cover the full frequency range of the bar pattern speed, from $9\lesssim\Omega_{\mathrm{pc}}\lesssim40$ rad Gyr$^{-1}$, implying that they contain $some$ information about all stages of the evolution. While higher PCs have their peak at higher frequencies, there is some power at low frequencies now clearly illustration the over-separation of the bar signal into many frequency components, matching Figures \ref{fig:LowLcorrelationmatrix} and \ref{fig:LowLPCs}. 

\begin{figure*}
\centering
\includegraphics[width=\hsize]{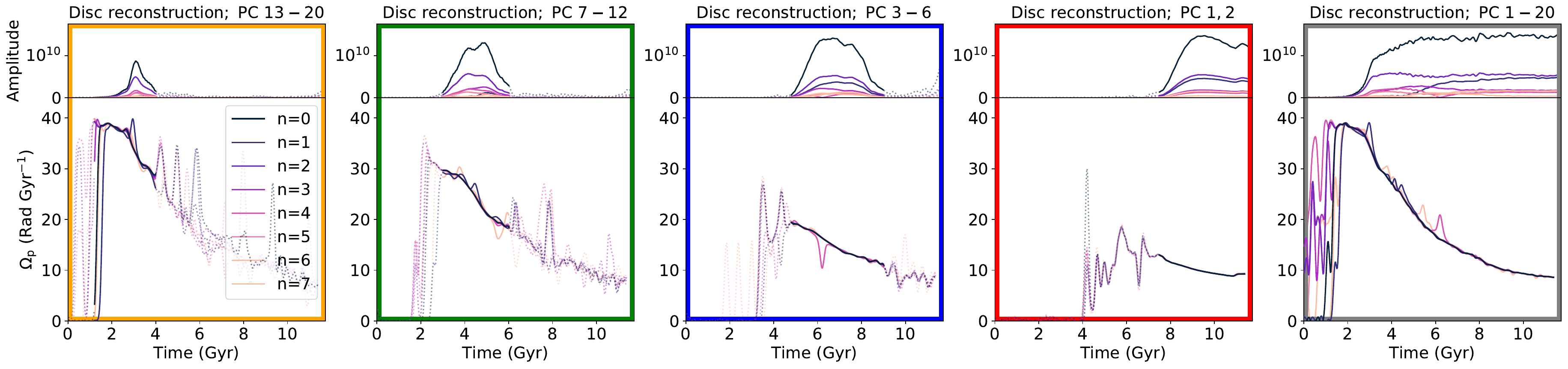}
\includegraphics[width=\hsize]{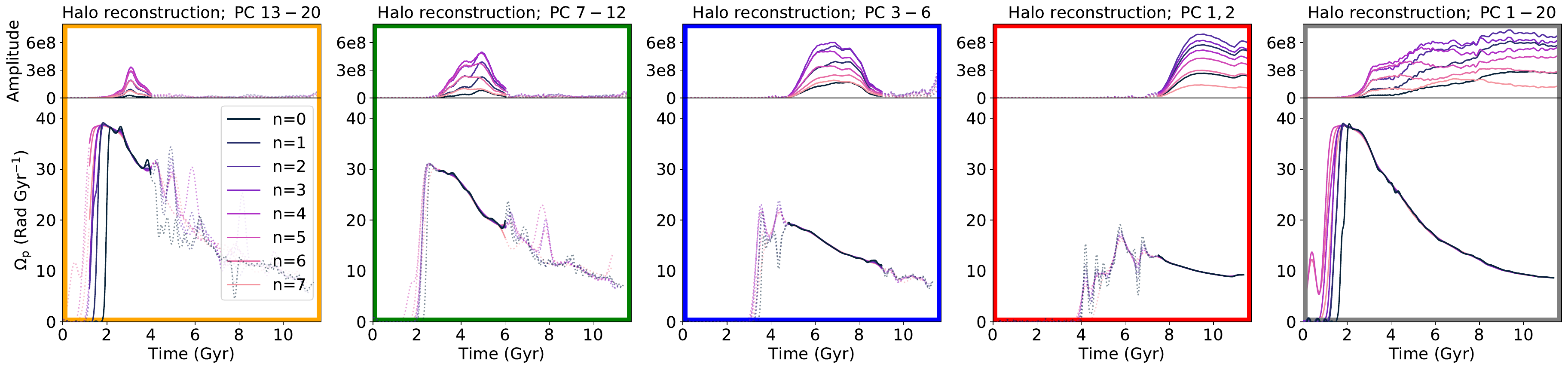}
\caption{Amplitude (upper sub-panels) and pattern speed (lower sub-panels) for the coefficient series reconstructed from given PC pairs (left to right) for the disc $m=2$ (upper row) and the halo $l=2, m=2$ (lower row), for $n=0-7$. Note that while the amplitude of the coefficients are close to zero the pattern speeds are non-existent or noisy (dotted lines), yet they become well defined and consistent with the frequency of the PC pair as the amplitude of the reconstruction grows (solid lines).}
\label{fig:LowLrecon_freqs}
\end{figure*}

\subsubsection{PC grouping}\label{sec:PC_grouping}

The PC pairs are not guaranteed to correspond to unique dynamical features, and some further grouping beyond the obvious pairs may be required (which requires human intervention) before reconstruction. The goal is to group the PCs such that we can interpret physically informative dynamics. In this instance, there are significant off-diagonal terms in the correlation matrix, such that there are many possible groupings.

In the absence of a clear separation in the correlation matrix or eigenvalues, we choose here to separate broadly by the epochs identified in the overall simulation evolution, i.e. $t\lesssim3$ Gyr (initial rapid growth phase: PC 13-20), $3\lesssim t\lesssim5$ Gyr (slow growth phase; PCs 7-12), $5\lesssim t\lesssim8$ Gyr (first steady state; PCs 3-6) and $t\gtrsim8$ Gyr (final steady state; PCs 1-2). We mark these groupings on Figure \ref{fig:LowLcorrelationmatrix} as Group 1 (PCs 1-2; red), group 2 (PCs 3-6; blue), Group 3 (PCs 7-12; green), Group 4 (PCs 13-20; orange) and Group 5 (`all significant PCs' 1-20; grey). 

\subsection{Reconstruction \& recovery of time evolving dynamical structures}\label{sec:mssa_apply}

\subsubsection{Coefficient series reconstruction}\label{sec:recon_coefs}
From these groupings, we then reconstruct the coefficient series from the contribution of the first twenty PCs. Figure \ref{fig:LowLreconCoefs} shows the reconstructed disc coefficients (upper row) and halo coefficients (lower row) for the first twenty PCs grouped as above (left four columns). The amplitude of the halo coefficients are approximately $5\%$ of that of the disc coefficients for a given group, yet the change in relative amplitude as a function of time is highly correlated between the disc and halo. This is because we are explicitly using mSSA to recover correlated signals in the disc and halo. The right hand column shows the reconstruction of all 20 PCs together. This matches the original $m=2$ coefficients for the disc and halo from Figure \ref{fig:coefs}, showing that the first 20 PCs are sufficient to reconstruct the bar and the halo response. 

We note here that the different $n$ orders across both the reconstructed disc and halo coefficients have different amplitudes, yet the same overall `shape' of the evolution of amplitude over time. We also note that while the disc $n=0$ coefficient is strongest, the halo coefficients are strongest at $n=2$ (representing scales of approximately 10 kpc) owing to the different radial extent of the halo and the disc. 

\begin{figure*}
\centering
\includegraphics[width=\hsize]{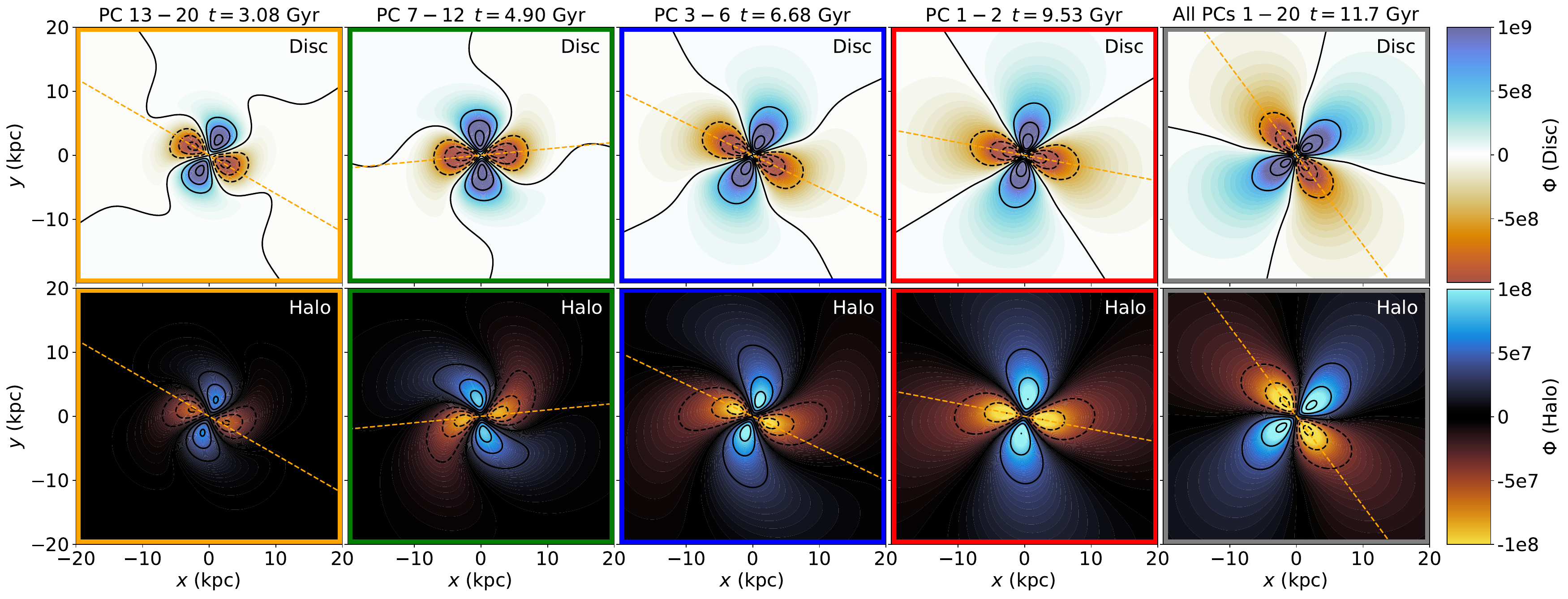}
\caption{$z=0$ slice of the quadrupole potential terms ($m=2$) in the stellar disc (upper row) and dark matter halo (lower row) reconstructed from the PC groups matching Fig's \ref{fig:LowLreconCoefs} and \ref{fig:LowLrecon_freqs}, plotted at the time of maximum PC amplitude. The dashed orange line marks the phase angle of the \textit{stellar} quadrupole on both rows. Note that the pattern rotation is clockwise}.
\label{fig:potential_wake}
\end{figure*}

\begin{figure*}
\centering
\includegraphics[width=\hsize]{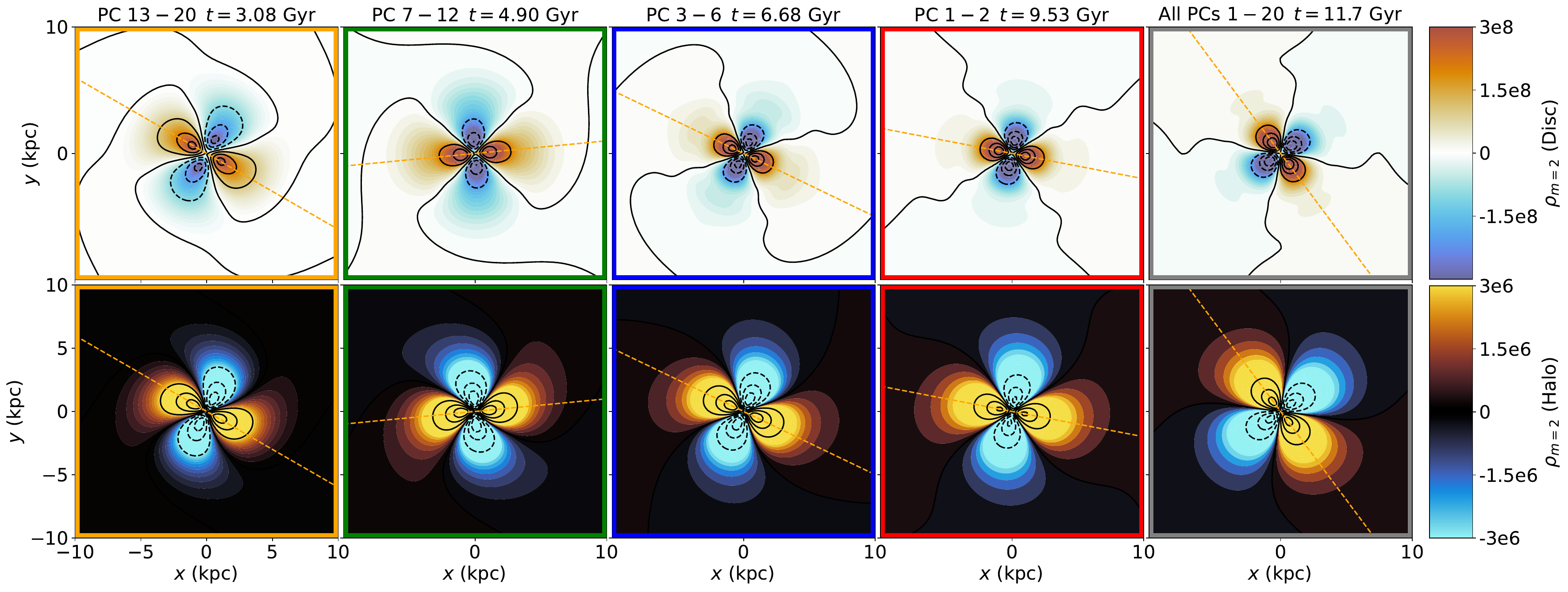}
\caption{Same as Figure \ref{fig:potential_wake} but for the density, $\rho$ (note the change in axis range).}
\label{fig:density_wake}
\end{figure*}

\begin{figure*}
\centering
\includegraphics[width=\hsize]{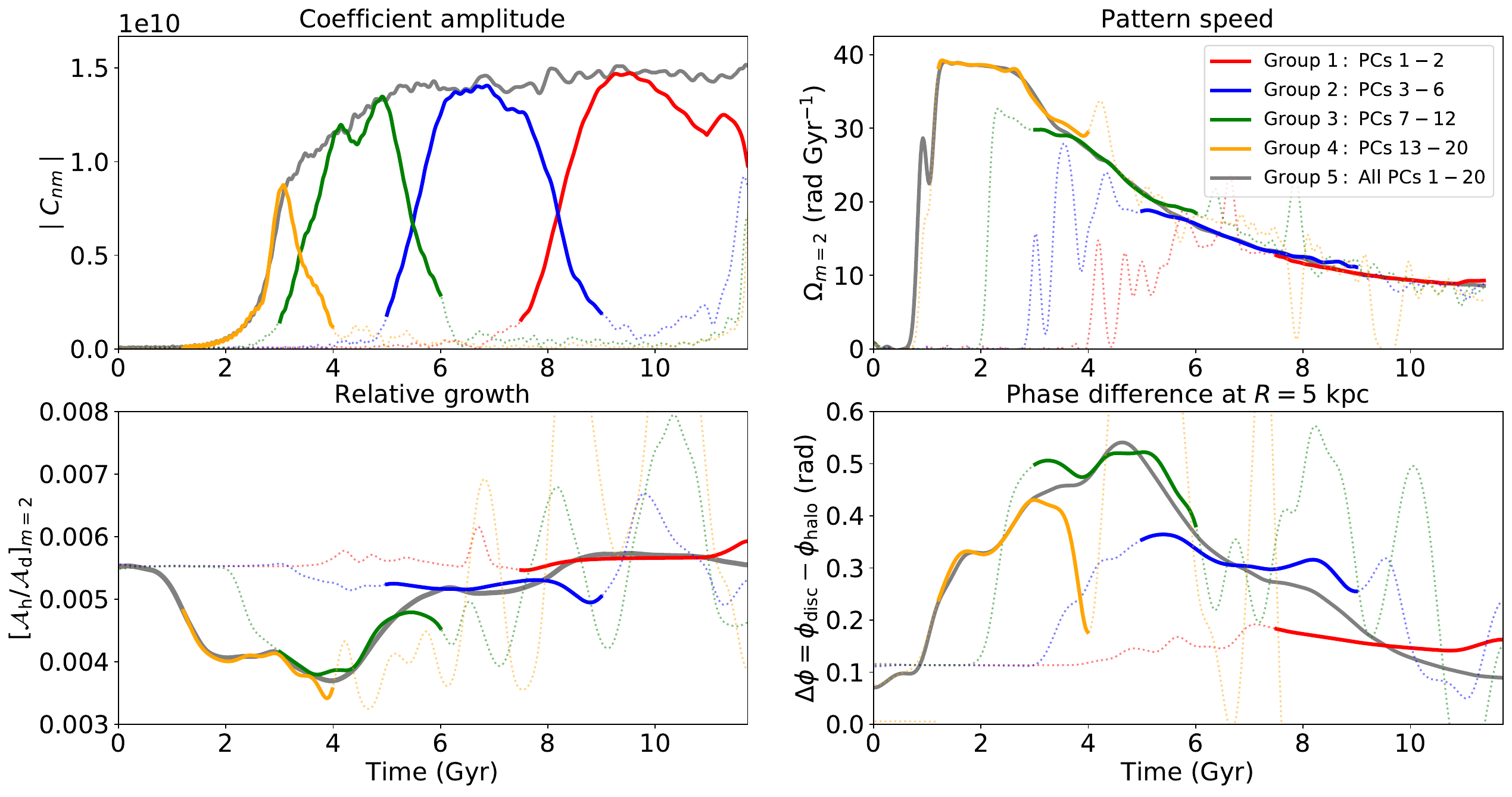}
\caption{\textbf{Upper left:} Amplitudes of the reconstructed $m=2$ disc coefficients as a function of time for the groups of PCs (colored lines) and PCs 1-20 (grey line). The lines are solid where the coefficients have significant amplitudes, and dotted where the amplitudes are negligible. This separation is kept consistent in the other three panels. \textbf{Upper right}: Pattern speed for the same groups. \textbf{Lower left:} The ratio of relative growth of the disc and halo modes over time. When the ratio decreases it implies the disc $m=2$ harmonic is growing faster than the halo, and when the ratio increases it implies the halo harmonic is growing faster than the disc. \textbf{Lower right:} The phase difference between the disc and halo $m=2$ harmonics.}
\label{fig:LowL_relative_growth}
\end{figure*}

The time evolution of the coefficients shows that the lower numbered PC pairs only appear to grow in amplitude at late times in the simulation, e.g. when the bar is strong and slow, as we expect from the frequencies of the PCs shown in Figure \ref{fig:LowLfreqs}. For example, PCs 1 \& 2 have a frequency peak of $\sim9.5$ \kmskpc (from Figure \ref{fig:LowLfreqs}), which is reached around $t\sim8$ Gyr in the pattern speed evolution shown in Figure \ref{fig:patternspeeds}, which is approximately when the coefficient series gains significant amplitude in the top left panels of Figure \ref{fig:LowLreconCoefs}. This behavior can be extrapolated to the other PC groups. 

As with the raw coefficients, we can again recover the pattern speed of the reconstructed coefficient series by using the real and imaginary components of the time series. Figure \ref{fig:LowLrecon_freqs} shows the amplitudes (upper sub-plot) and frequencies (lower sub-plot) over time of the coefficients reconstructed from the PC groups for the disc (upper row) and the halo (lower row). The different groups show different sections of the overall pattern speed evolution, with the combined PCs 1-20 in the right hand column again reproducing the overall bar slow down. The groups show well-measured pattern speeds when the amplitude of the group is large (solid lines), yet they are either noisy (owing to over-separation, again as evidenced by the correlation matrix in Figure \ref{fig:LowLcorrelationmatrix}) or consistent with zero (before the bar reaches that stage of evolution) outside these time periods (dotted lines). 

At early times we see the $m=2$ mode for the halo components is dominant at small radial scales. While the large scale disc $m=2$ is always largest even at early times, in both the disc and halo the dominant power evolves to be at larger scales with a slower pattern speed. The decomposition into PCs quantifies how the bar in this simulation slows as it grows. It is again clear that the growth and pattern speed evolution of the bar and halo are closely linked, and that mSSA can separate the early time growth phase and late time steady state of these coupled dynamical modes. 

\subsubsection{Field reconstruction}\label{sec:field_recon}

From the reconstructed coefficients, we can now reconstruct various fields in combination with the bases used to originally calculate the coefficients.

Figure's \ref{fig:potential_wake} and \ref{fig:density_wake} show the reconstructed $m=2$ potential and density fields respectively, for the disc (upper rows) and dark halo (lower rows) for the PC groupings as explored above, plotted at the timestep where each group has maximum amplitude such that time increases left to right, albeit not linearly. For all groups we see clearly that the $m=2$ bar component of the stellar disc has a corresponding bar-like overdensity and trailing wake in the dark matter potential and density. This is expected from previous studies \citep[e.g.][]{Petersen+16_shadowbar}, but the mSSA machinery explicitly tells us that they are dynamically coupled and share evolution over these timescales. We see clearly the growth of the central quadrupole (the trapped ``shadow bar'') and the untrapped trailing dark matter wake which grows in both physical extent and amplitude along with the growth of the stellar bar. 

The dashed orange line marks the orientation of the stellar bar at the given timestep, also plotted on the dark halo response. From this, we see that the central part of the dark $m=2$ pattern is aligned with the stellar bar. The trapped `shadow bar' is always aligned with the stellar bar by definition; it is the same ILR orbits as the stellar orbits in the same overall potential. The shadow bar grows before the wake because it grows at the same rate as the stellar bar, at least at the outset, as was already shown in Figure \ref{fig:bar_growth}.

There then extends an untrapped response which manifests as a trailing wake in the dark matter. This lag of the wake behind the bar provides a torque which slows the bar over time \citep[e.g.][]{Hernquist+Weinberg92}. Figures \ref{fig:potential_wake} and \ref{fig:density_wake} show that the difference in phase between the stellar bar and the dark matter wake decreases over time as explored in the next section. This occurs because at early times, the resonant interactions between disc and halo result in angular momentum transfer which appear as a lagged wake. At late times, the wake becomes non-resonant, feeling the bar but not accepting angular momentum from the bar. With no symmetry breaking by resonances, the untrapped response becomes aligned with the bar.

\subsection{Identifying coupling \& causation}\label{sec:mssa_candc}

We can use this representation to explicitly demonstrate coupling of the disc and halo, and the causal relationship.

The upper left panel of Figure \ref{fig:LowL_relative_growth} shows simply the amplitudes of the disc (solid) and halo (dashed) coefficients reconstructed from the PC groups determined above as a reminder and illustration of how groups 1-4 (colored lines) represent different stages of bar evolution and that they combine to reproduce the full signal (grey). The disc and halo coefficients for these PCs are visually very similar and thus only the disc is shown for simplicity. The lines are dotted where the amplitude is not significant, which is maintained across the other panels.

\subsubsection{Pattern speed evolution}

The upper right panel of Figure \ref{fig:LowL_relative_growth} shows the pattern speed evolution of the same PC groups, with solid lines where the amplitude from the upper left panel is significant, and dotted where it is not. Again, the combination of groups reproduces the overall pattern speed evolution of the bar matching Figure \ref{fig:patternspeeds}.

We can also measure the slowdown of the bar from this curve. While we note that we could also have done this from the original $m=2$ coefficient curve of Figure \ref{fig:patternspeeds}, the mSSA extracted bar PCs should be free from other $m=2$ patterns which could bias the measurement. We find that a power law of form $t^{-1}$ is a reasonable approximation over the $\sim9$ Gyr of deceleration from $3<t\lesssim12$ \citep[matching for example the form assumed in][in their study of bar deceleration]{CFS19}. However, the power law of temporally local segments of the evolution can differ from this. For example, the late stage evolution for $t>9$ Gyr is well fit by a power law of the form $t^{-0.7}$. This is likely reflective of the fact that the bar is not expected to slow completely to $\Omega_\mathrm{b}=0$ \kmskpc (i.e. fully stop), and rather that the pattern speed should proceed asymptotically to some small positive value as the resonant pathways for angular momentum exchange become saturated.

\subsubsection{Relative growth rates}

We know from theory and earlier work that the shadow bar and wake are formed in response to the stellar bar \citep[e.g.][]{Hernquist+Weinberg92}, and that the dark matter wake in turn exerts a torque on the bar which causes it to slow down. This can be shown using the change in relative amplitudes of the reconstructed coefficient series using the equation
\begin{equation}
[\mathcal{A}_{\mathrm{h}}/\mathcal{A}_{\mathrm{\mathrm{d}}}]_{m=2}=\sum_{n=0}^{n_\mathrm{max}}\frac{{M_d\mid C_{\mathrm{h},l=2,m=2,n}\mid}}{{M_{\mathrm{h}}\mid C_{\mathrm{d},m=2,n}\mid}},
\end{equation}
i.e. the ratio between the sum of the amplitudes of the $m=2$ coefficient series for the disc and halo normalised by relative disc and halo mass. 

The lower left panel of Figure \ref{fig:LowL_relative_growth} shows this relation over time for the four PC groups (colored lines) and the sum of the first twenty PCs in grey. The lines are dotted when the group has close to zero amplitude matching the top left panel, both before the group exists (when the ratio is unchanged) and after it decays (when the ratio becomes noisy).

The solid grey line shows that for all PCs the ratio is constant for $\sim1$ Gyr during the exponential growth phase while the stellar disc instability and corresponding `shadow bar' grow together as seen in the upper panel of Figure \ref{fig:bar_growth}. From $1.2\lesssim t\lesssim2$ Gyr the ratio drops, showing that the disc $m=2$ instability is growing much faster than the halo. From $t\sim2$ Gyr the ratio stabilises, and the disc and halo $m=2$ instability grow at approximately the same rate until $t\sim4$ Gyr. From $t\sim4$ to $t\sim6$ Gyr the halo $m=2$ grows faster in the halo than in the disc, until it again stabilises from around $6<t<8$ before growing once more to a final stable state at $t\sim8$ Gyr.

This behaviour is consistent with the evolution of the raw coefficients as shown in Figure \ref{fig:coefs}, and the pattern speed curve in Figure \ref{fig:patternspeeds}. For example, from $t\lesssim2$ Gyr a weak $m=2$ instability with a fixed pattern speed grows in the disc. From $2\lesssim t\lesssim4$ Gyr the halo responds, and they grow together significantly enhancing the disc $m=2$, while the pattern speed of the disc and halo $m=2$ begin to slow down together. From $4\lesssim t\lesssim8$ the amplitude of the halo $m=2$ mode experiences periods of growth, while the disc does not. Finally, from $t \gtrsim8$ Gyr both disc and halo are relatively stable.

The groups clearly capture different stages of the bar evolution, e.g. Group 4 (yellow) is capturing the initial growth of $m=2$ disc instability and the halo response. Group 3 (green) and Group 2 (blue) are capturing the intermediate stages of evolution when the halo response is growing faster than the stellar bar growth, and Group 1 (red) is capturing the late dynamics where both the disc and halo $m=2$ has reached a (close to) steady state. 

\subsubsection{Phase lag and torque}
We can also quantify the evolution of the phase difference between the stellar bar quadrupole and the dark matter halo response. We expect the halo wake to lag behind the bar, creating torque which then slows the bar down.

The lower right panel of Figure \ref{fig:LowL_relative_growth} shows the phase difference between the angle of the quadrupole in the stellar disc and the dark halo at $R=5$ kpc. We see that at all times the phase difference is positive, i.e. the dark matter response is trailing the bar, as expected. At early times as the bar is growing (Group 5; orange) the phase difference increases rapidly as the stellar bar grows faster than the halo response. Then, this phase difference stabilises (in Group 4; green) as the overall growth rate slows (see upper left panel) and the halo response grows faster than the stellar quadrupole (see lower left panel). Then for the later evolution Groups 1 (red) and 2 (blue), we see a decrease in the phase lag along with the pattern speed, and the absolute and relative growth panels show close to a steady state. In general we note that the larger the phase difference, the larger the torque, and the higher the rate of slowdown of the bar, as reflected in the evolution of the pattern speed in the upper right panel.

We also note that from $5\lesssim t\lesssim12$ Gyr, while Groups 1 (red) and 2 (blue) show a decrease in the phase lag with time, there is an apparent `bump' in the evolution around $t\approx8$ Gyr, which corresponds to the slight change in relative growth rate in the lower left panel, and the slight increase in the overall coefficient amplitude seen in the upper left panel (or more clearly in Figure \ref{fig:coefs}), suggesting a real change in the dynamics of the bar-halo coupling.

Figure \ref{fig:lag_fit} shows two power law fits to the declining region of the phase lag curve from the lower right panel of Figure \ref{fig:LowL_relative_growth}. Indeed we find that the segments before and after $t\approx8$ Gyr are best fit with power laws of the form $t^{-1.8}$ and $t^{-3.3}$ respectively.

However, it is known that the dynamics of the interaction and the transfer of angular momentum depend on the profile and velocity dispersion of the disc and halo. As such, a more thorough exploration of bar-halo coupling as a function of such parameters will be the focus of a future work. For now, we note how the mSSA analysis both nicely illustrates the `big picture' dynamical evolution of the bar and the response in the dark halo, but also allows us to quantify the evolution of specific regimes of the bar-halo coupling relatively straightforwardly, while beforehand it is difficult to extract such subtle signals from particle phase space, let alone make detailed measurements of coupled growth rates and phase lags over dynamically distinct temporal regimes.

\begin{figure}
\centering
\includegraphics[width=\hsize]{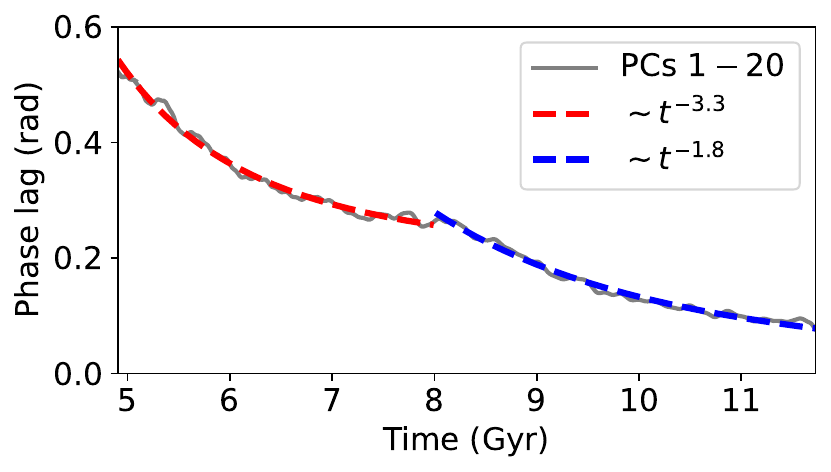}
\caption{Power law fits to the two regimes of decrease in the angle of the lag between the stellar and dark matter quadrupole.}
\label{fig:lag_fit}
\end{figure}

\section{Summary \& conclusions} \label{sec:summary}

In this work, we investigate the coupled dynamical evolution of a stellar bar and dark matter halo using a high-resolution ($\sim10^9$ particles), isolated disc galaxy simulation. We use an adaptive basis function expansion (with the EXP framework; \citealt{exp2025JOSS}) to represent and quantify the evolution of the mass distribution of each component (bulge, disk, and halo). We then use multichannel Singular Spectral Analysis (mSSA) to study the physical connection between the bar’s growth and slowdown and the corresponding dark matter response. Our conclusions are as follows:
\begin{itemize}
\item \textbf{A clear coupling between bar and dark matter halo:} The stellar bar induces a dynamically-coupled response in the dark matter distribution. This manifests as both an inner \textit{shadow bar} aligned with the stellar bar \citep[as seen e.g. in][]{Petersen+16_shadowbar} and an extended dark matter wake that trails (in phase) the stellar bar. This reaffirms that the dark matter can actively exchange angular momentum with the disc, which further couples the dynamical evolution of these Galactic components.
\item \textbf{Distinct phases in the bar evolution:} The bar--halo system evolves through a few identifiable regimes: rapid exponential growth of the bar, intermediate adjustment, and a long-term quasi-steady phase. Each of these phases is characterized by unique growth rates and phase relationships between the disc and halo quadrupoles.
\item \textbf{Bar slowdown and angular momentum exchange:} For the bulk of its evolution, the pattern speed of the stellar bar decreases approximately as $t^{-1}$, consistent with expectations for angular momentum transfer from the bar to the halo because of a phase lag in the dark matter response. This phase lag --- between the stellar and dark matter components --- decreases over time, which causes the late-time evolution of the pattern speed to evolve more slowly ($\approx t^{-0.7}$).
\item \textbf{Demonstration of mSSA as a tool for dynamical discovery:} The combination of EXP basis function expansion coefficients and mSSA has enabled us to successfully isolate and reconstruct complex, coupled dynamical modes from a high-resolution $N$-body simulation. This framework provides a non-parametric scheme for quantifying the detailed evolution of modes --- their relative growth, phase, and dynamical exchange --- which are not easily distinguishable by eye, and which are often more difficult or require more hand tuning with conventional analysis methods.
\end{itemize}
This work provides a demonstration of a new analysis framework for disentangling complex dynamical processes in galaxies. However, this analysis was performed on a single idealized, isolated system with a kinematically hot disc. A follow up systematic study of bar-halo coupling as a function of galactic disc and halo parameters is required before drawing generalisable conclusions. Applying these methods to simulations in more realistic, cosmological environments, and in isolated systems with different halo profiles, velocity dispersions, and mass components, will help establish a more comprehensive picture of the evolution of galactic structure across cosmic time.

\vspace{-4pt}

\section*{Data availability}
The simulation is available on reasonable request to the lead author. The Basis Function Expansion based simulation and analysis framework \texttt{exp}\ \citep{exp2025JOSS} and associated tutorials can be found online at \url{https://github.com/EXP-code}.

\vspace{-4pt}

\section*{Acknowledgements}
We thank the anonymous referee for their constructive report. JASH acknowledges the support of a UKRI Ernest Rutherford Fellowship ST/Z510245/1. MSP is supported by a UKRI Stephen Hawking Fellowship. KVJ is supported by Simons Foundation grant 1018465. MB acknowledge the grants PID2021-125451NA-I00 and CNS2022-135232 funded by MICIU/AEI/10.13039/501100011033 and by ``ERDF A way of making Europe'', by the ``European Union'' and by the ``European Union Next Generation EU/PRTR''. The B-BFE collaboration gratefully acknowledges the support of Simons Foundation for their collaboration meetings.  KJD and S\'OH acknowledge support from the Heising Simons Foundation grant \# 2022-3927. They also respectfully acknowledge that the University of Arizona is home to the O'odham and the Yaqui. We respect and honor the ancestral caretakers of the land, from time immemorial until now, and into the future. The majority of this work was performed on the Flatiron Institute `Rusty' Iron Cluster, and supported by their Scientific Computing Core. This work makes extensive use of the Basis Function Expansion based simulation and analysis framework \texttt{exp} \citep{exp2025JOSS}. We have used the scientific libraries \texttt{numpy}\ \citep{numpy}, \texttt{SciPy}\ \citep{scipy}, \texttt{Agama}\ \citep{agama} and \texttt{CMasher}\ \citep{cmasher} in this work.

\vspace{-4pt}
\bibliographystyle{mn2e}
\bibliography{ref2}

\label{lastpage}
\end{document}